\documentclass[aps,twocolumn,groupedaddress,prb]{revtex4}

\usepackage{graphicx}
\usepackage{dcolumn}
\usepackage{bm}
\usepackage{amsmath}

\begin{document}


\title{\textit{Ab initio} theory of gate induced gaps in graphene bilayers}

\author{Hongki Min$^1$}
\email{hongki@physics.utexas.edu}
\author{Bhagawan Sahu$^2$}
\author{Sanjay K. Banerjee$^2$}
\author{A. H. MacDonald$^1$}
\affiliation{
$^1$Department of Physics, The University of Texas at Austin, Austin Texas 78712, USA\\
$^2$Microelectronics Research Center, The University of Texas at Austin, Austin Texas 78758, USA
}

\date{December 10, 2006}

\begin{abstract}
We study the gate-voltage induced gap that occurs in graphene bilayers
using \textit{ab initio} density functional theory.  Our calculations
confirm the qualitative picture suggested by phenomenological tight-binding
and continuum models.  We discuss enhanced screening of the
external interlayer potential at small gate voltages,
which is more pronounced in the \textit{ab initio} calculations,
and quantify the role of crystalline inhomogeneity using a tight-binding model
self-consistent Hartree calculation.
\end{abstract}

\maketitle

\section{Introduction}
Recently, ultrathin graphite films including monolayers\cite{novoselov2004,novoselov2005a,zhang2005a}
and bilayers \cite{novoselov2006,ohta2006} have attracted considerable attention because of their novel properties.
In single-layer graphene, the $A$ sublattice to $B$ sublattice hopping amplitude vanishes at two inequivalent
points $K$ and $K'$ on the edge of the honeycomb lattice Brillouin zone (BZ); away from these points, the hopping amplitude grows
linearly with the wave vector and has a phase which winds along with the orientation of the wave vector
measured from the high-symmetry points.  The band structure of an isolated graphene layer is therefore described at low energies by
a two-dimensional massless Dirac equation with linear dispersion; this property
gives rise to a half integer quantum Hall effect\cite{novoselov2005b,zhang2005b}
and to a quantized spin Hall effect\cite{kane2005,sinitsyn2006} and dominates the low-energy physics.

\begin{figure}[h]
\scalebox{0.45}{\includegraphics{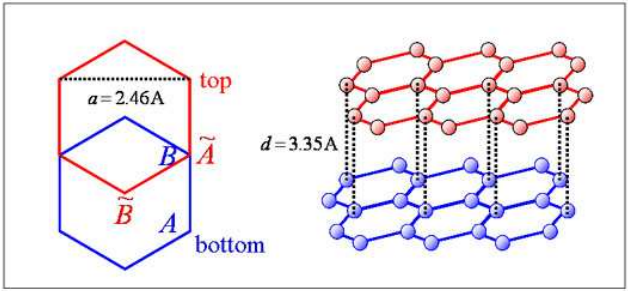}}
\caption{(Color online) Structure of a graphene bilayer with honeycomb lattice constant $a=2.46$ $\rm \AA$
and interlayer separation $d=3.35$ $\rm \AA$.}
\label{fig:bilayer}
\end{figure}

\begin{figure}[h]
\scalebox{0.45}{\includegraphics{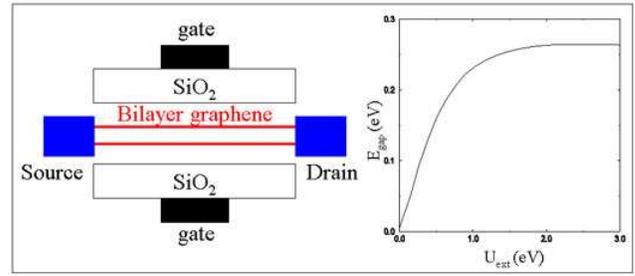}}
\caption{(Color online) Schematic illustration of a circuit with a
bilayer graphene channel sensitive to an external gate voltage.  The graphene
channel is separated from the front and back gates by a $\rm{SiO_2}$ layer.
The channel resistance change will be rapid and large when the graphene channel is
undoped and isolated from the gate electrodes, as illustrated here.
In this case, the total charge density in the bilayer system is fixed and the
chemical potential lies in the gap opened by the gate voltage.
This geometry could also be used to capacitively probe
the correlation physics of the isolated bilayer system, as discussed in the text.
}
\label{fig:device_gap}
\end{figure}

In bilayer graphene, the Bernal ($\tilde{A}$-$B$) stacking
illustrated in Fig. \ref{fig:bilayer} modifies this electronic structure in
an interesting way.\cite{novoselov2006,latil2006}  At $K$ and $K'$, the states localized at
the $\tilde A$ and $B$ sites, are repelled from zero energy by interlayer tunneling;
only states localized at $A$ and $\tilde B$ are present at zero energy.
When tunneling is included, the $A$ to $\tilde B$ hopping is a second-order
process via a virtual bonding or antibonding state at $\tilde A$ and $B$.
The chirality of the low-energy bands is therefore doubled.
Most intriguingly, an external potential which induces a difference between the $A$ and $\tilde B$ site
energies will open up a gap\cite{mccann2006a,mccann2006b} in the spectrum.
Band gaps controlled by applying a gate bias
have been studied experimentally using angle-resolved photoemission spectroscopy\cite{ohta2006}
and Shubnikov-de Haas analysis\cite{castro2006} of magnetotransport.
This unique property of bilayer graphene has created considerable interest
in part because it suggests the possibility of switching
the conductance of a graphene bilayer channel over a wide range at a speed which is limited
by gate-voltage switching, as illustrated schematically in Fig. \ref{fig:device_gap}.

In this paper, we report on an \textit{ab initio} density functional theory (DFT) study of
the influence of an external potential difference between the layers on the electronic
structure of a graphene bilayer.  We compare our results with the phenomenological
tight-binding and continuum model Schr\"{o}dinger-Poisson calculations used in
previous theoretical analyses.{\rm \cite{mccann2006b}}  DFT predicts, in agreement with these works, that the
external potential difference is strongly screened with a maximum energy gap value of $\sim$0.3 eV.
There are, however, quantitative differences.  In particular, the enhanced screening
which occurs for weak external potentials is stronger in the DFT calculations.
In an effort to improve the quantitative agreement, we have estimated the
influence of crystalline inhomogeneity in a tight-binding model self-consistent Hartree calculation.
This effect strengthens intralayer Coulomb interactions because the charge is
spatially bunched, and therefore increases screening in a Hartree calculation,
but does not fully account for differences between the two calculations.

\section{\textit{Ab initio} Density Functional Theory Calculations}
We have performed \textit{ab initio} DFT calculations\cite{kohn1965}
for an isolated graphene bilayer under a perpendicular external electric field
using an all-electron linearized augmented plane wave plus local-orbital method
incorporated in WIEN2K.\cite{blaha2001}
We used the generalized-gradient approximation\cite{perdew1996} for the exchange and correlation potential.

\subsection{External electric fields}
To investigate the influence of an external electric field on a graphene bilayer,
a periodic zigzag potential was applied along the $z$ direction,
perpendicular to the graphene planes, in a supercell.\cite{stahn2001}
The bilayer was placed at the center of the constant external electric field region
and the size of the supercell was set to a large value ($\sim$16 ${\rm\AA}$)
to minimize the interaction between bilayers in neighboring supercells.
In order to resolve the small gaps produced by small external fields we performed BZ
sums using a relatively large number of $\bf{k}$ points ($\sim$800) per irreducible wedge (5000 $\bf{k}$ points in the whole BZ).
Total energies were convergent to within 0.0001 Ry.

\begin{figure}[h]
\scalebox{0.45}{\includegraphics{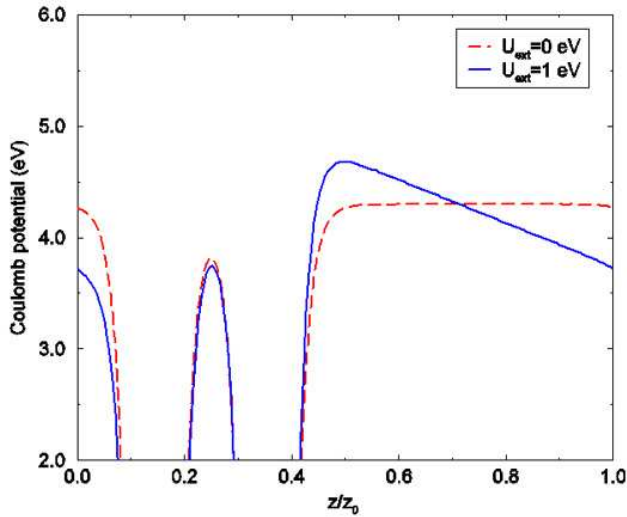}}
\caption{(Color online) An averaged Coulomb potential of a cross section
vs $z$ for total external potential $U_{ext}=0$ eV and $U_{ext}=1$ eV.
Here, $z_0$ is the superlattice period and the cross section
was chosen to include equal number of atoms in each layer.}
\label{fig:potential}
\end{figure}
Figure \ref{fig:potential} shows the Coulomb potential relative to the Fermi energy,
laterally averaged along a line in the $x$-$y$ plane that includes an equal number of atoms in each layer,
as a function of the $z$ coordinate.
The potential includes the Hartree electron-electron potential and the electron-ion interaction but not the external electric
field potential or the exchange-correlation potential.
The bilayer is centered around $z/z_0=0.25$, where $z_0$ is the superlattice period.
(In the discussion below, we define the external potential energy $U_{ext}$
as $U_{ext}=e E_{z,ext} d$, where $E_{z,ext}$ is an external electric field
along the $z$ direction and $d$ is the interlayer separation of bilayer graphene which
we take to be 3.35 $\rm{\AA}$.)
In the absence of an electric field (dashed line), the Coulomb potential is flat
in the vacuum region and the energy difference between the vacuum and the Fermi energy
gives estimates of the work function of bilayer graphene to be $\sim$4.3 eV.
In the presence of an electric field (solid line), charge transfer between the layers induces a potential
which cancels the external potential in the vacuum region.
The difference between the Coulomb energies of the two layers in the presence of
an external electric field is closely related to the gate-voltage induced energy gap.

\begin{figure}[h]
\scalebox{0.45}{\includegraphics{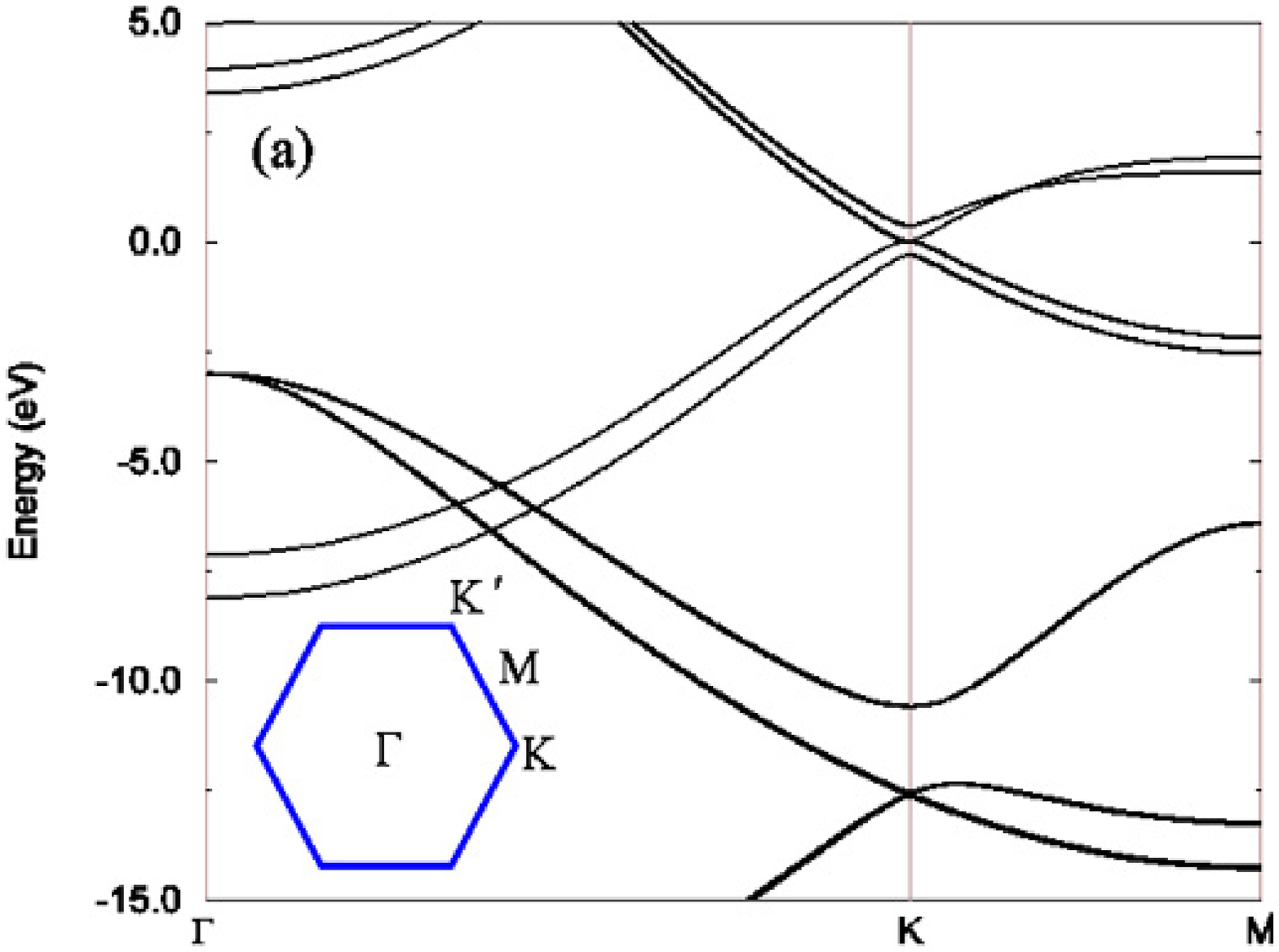}}
\scalebox{0.46}{\includegraphics{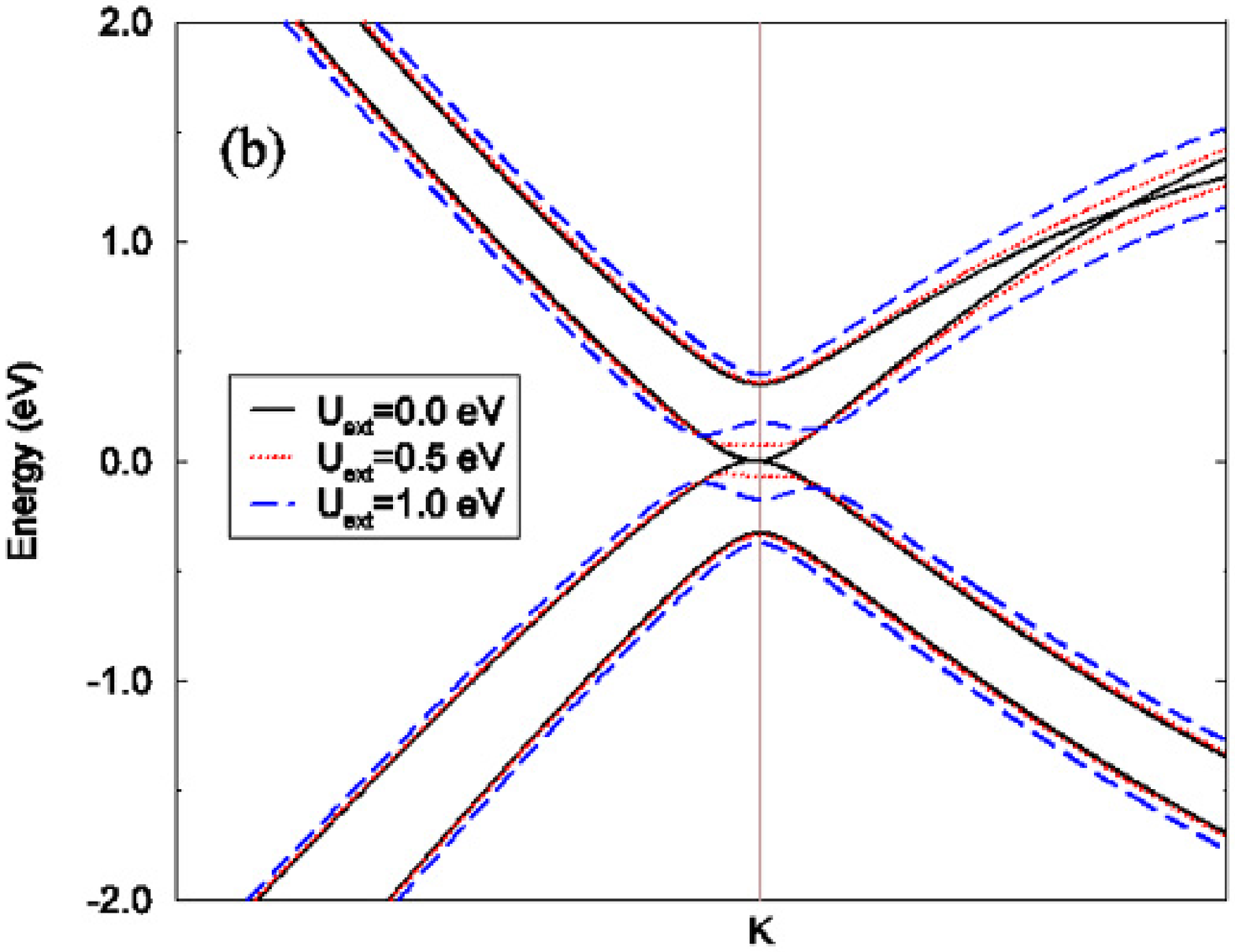}}
\caption{(Color online) (a) Bilayer graphene band structure in the absence of an external electric field.
(b) Bilayer graphene band structure near the $K$ point for $U_{ext}$=0, 0.5, and 1 eV.}
\label{fig:band}
\end{figure}

\subsection{Energy bands}
Figure \ref{fig:band}(a) shows the DFT energy band structure of bilayer graphene
in the absence of an applied external electric field.
When $U_{ext}=0$ eV, the low-energy band dispersion is nearly parabolic at two inequivalent corners, $K$ and $K'$,
of the hexagonal BZ, as predicted by the $\pi$-orbital tight-binding and continuum model phenomenologies.\cite{mccann2006a,mccann2006b}
The valence and conduction bands meet at the Fermi level.

In the absence of an external electric field, bilayer graphene, like single-layer graphene, is a zero-gap semiconductor.
At finite $U_{ext}$, however, the low-energy bands near the $K$ or $K'$ point split,
as explained in the Introduction. Therefore, gated graphene bilayer systems are gate-voltage tunable
narrow gap semiconductors [Fig. \ref{fig:band}(b)].  This property is unique, to our knowledge.
It is worth noting that in the presence of an external electric field,
the true energy gap does not occur at the $K$ or $K'$ point but slightly away from it.
The low-energy spectrum develops a \textit{Mexican hat} structure as the strength of the external electric field increases.
This property is also captured by phenomenological models of graphene bilayers.\cite{mccann2006b}

\subsection{Evolution of tight-binding model parameters with $U_{ext}$}

\begin{figure}[h]
\scalebox{0.45}{\includegraphics{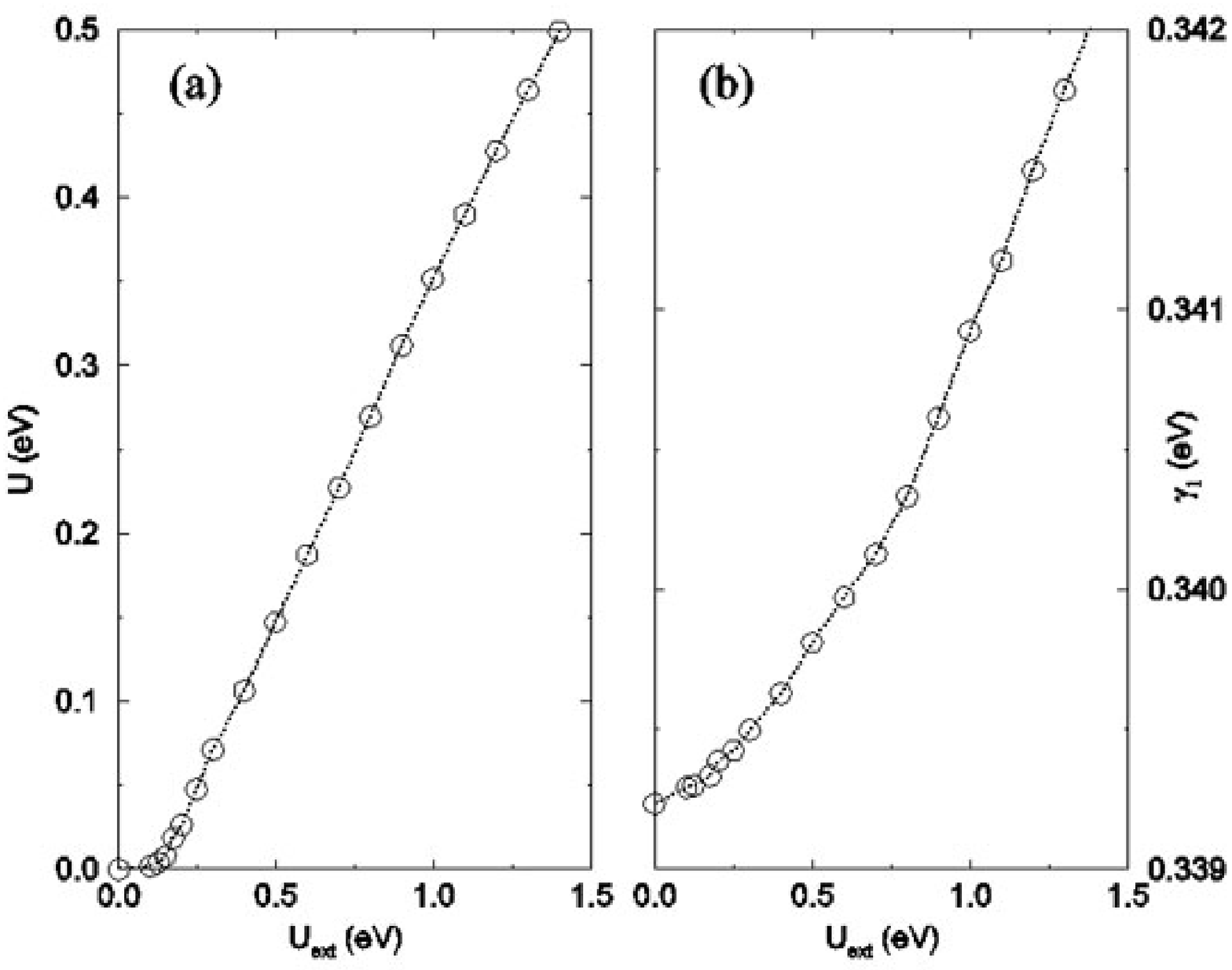}}
\scalebox{0.45}{\includegraphics{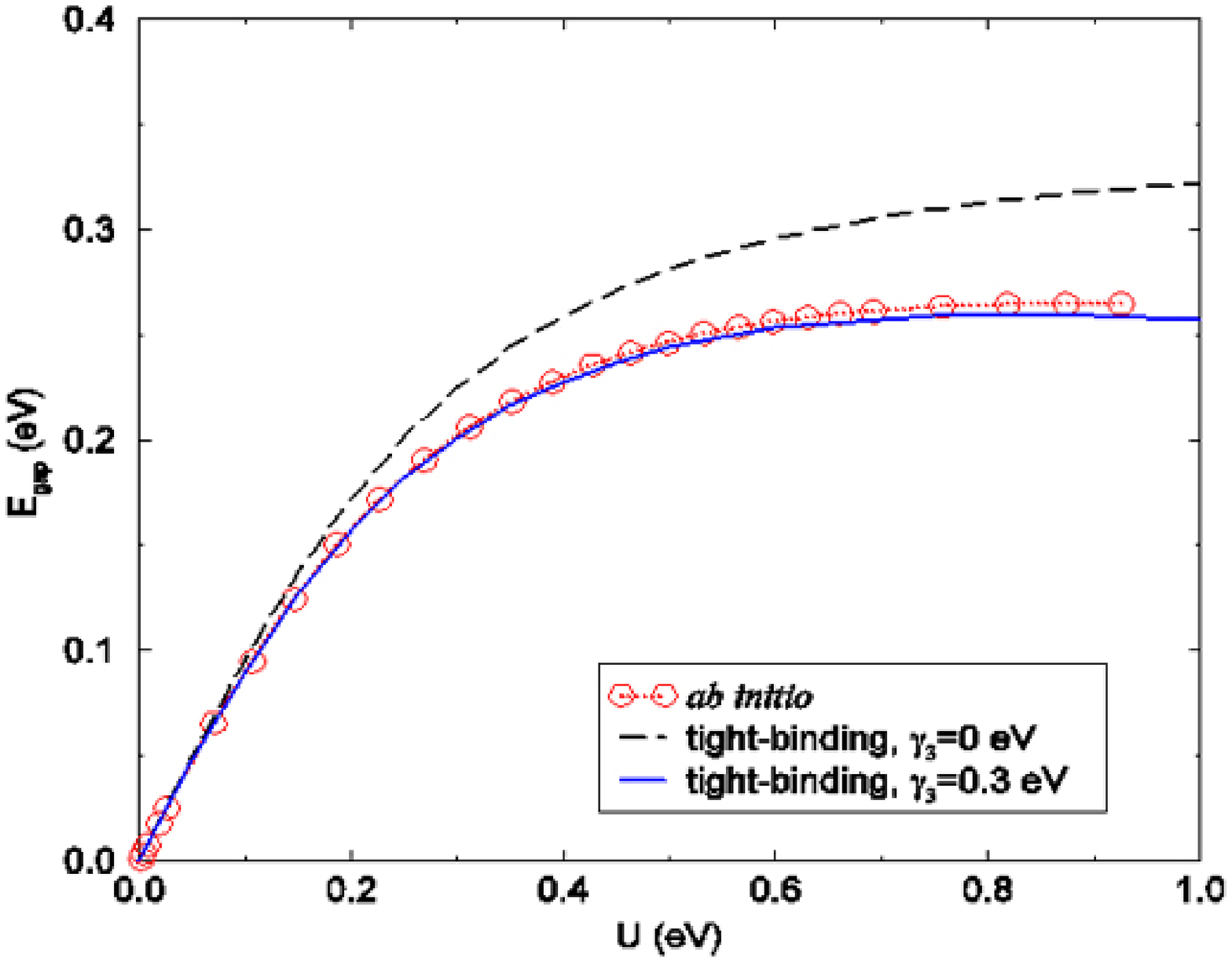}}
\caption{(Color online)
(a) Evolution of the graphene bilayer screened on-site energy difference $U$, extracted
from the \textit{ab initio} DFT bands as explained in the text, with the external potential $U_{ext}$.
The external potential is strongly screened.
(b) Evolution of the interlayer tunneling amplitude $\gamma_1$ with $U_{ext}$.
(c) Comparison of band gap as a function of the on-site energy difference $U$
obtained from the \textit{ab initio} DFT calculations (open circles) with the tight-binding result
for $\gamma_3=0$ (dashed line) and $\gamma_3=0.3$ eV (solid line).
}
\label{fig:parameter}
\end{figure}

Figure \ref{fig:parameter} illustrates DFT predictions for the evolution of tight-binding parameters
with the applied external potential.  The tight-binding model expression for the four low-energy
band eigenvalues at the $K$ and $K'$ points is $E_{K/K'}=\pm U/2$, $\pm \sqrt{\gamma_1^2+U^2/4}$,\cite{mccann2006a}
where $U$ is the interlayer energy difference and $\gamma_1$ is the interlayer tunneling amplitude.
(As we discuss below, this expression should, strictly speaking, be slightly modified in the presence
of an external potential, but it still provides a convenient way of characterizing DFT
predictions for the low-energy bands.)  The values of $U$ and $\gamma_1$ plotted in
Fig. \ref{fig:parameter} represent this interpretation of the four lowest-energy DFT eigenvalues
and clearly reflect substantial screening of the external interlayer potential by the
Hartree potential plotted in Fig. \ref{fig:potential}.
The interlayer coupling $\gamma_1$ increases monotonically as the external potential increases.
The rate of increase is, however, ten times smaller than estimated in a
recent experimental study\cite{ohta2006} of a doped bilayer systems, possibly
suggesting significant differences between doped and undoped systems.
The intralayer nearest-neighbor $\pi$-electron hopping amplitude $\gamma_0$
and the interlayer $A$-$\tilde{B}$ coupling $\gamma_3$
were fitted to reproduce the band dispersion around the $K/K'$ points at low energies.
We find $\gamma_0\approx 2.6$ eV and $\gamma_3\approx 0.3$ eV, nearly independent of the external electric field.
This value for $\gamma_0$ corresponds to an in-plane velocity
$v={\sqrt{3}\over 2}{a \gamma_0\over \hbar} \approx 8.4\times 10^5$ m/s,
where the lattice constant $a=2.46$ ${\rm \AA}$.

Figure \ref{fig:parameter}(c) compares the relationship between the on-site energy difference $U$ extracted
from the DFT calculations and the energy gap with the corresponding relationship in the tight-binding
model.  Note that the gap does not increase indefinitely with $U$ but saturates at $\sim$0.3 eV
due to the \textit{Mexican hat} structure shown in the bands illustrated in Fig.\ref{fig:band}.
For $\gamma_3=0$, we can estimate the approximate energy gap from the low-energy approximation of the tight-binding model
given by $E_{gap}\approx |U|\gamma_1/ \sqrt{\gamma_1^2+U^2}$,
where $E_{gap}$ approaches $\gamma_1\approx 0.34$ eV as $U$ increases.\cite{mccann2006b}
For $\gamma_3\approx 0.3$ eV, however, $E_{gap}$ is reduced from that of $\gamma_3=0$
and matches well with the DFT results.
A nonzero value for $\gamma_3$ has a noticeable quantitative influence on the bands.
This agreement confirms (unsurprisingly)
that the tight-binding model captures the character of the low-energy bands in bilayer graphene.
The most interesting physics is in the relationship between $U$ and $U_{ext}$, which we now examine
more closely.

\section{Screening Theories}

\subsection{Continuum Hartree potential models}

The screening of the external potential has been examined previously for both
doped and undoped bilayers using phenomenological approaches combined with
the Poisson equation.\cite{mccann2006b}
This type of analysis provides a good reference point for interpreting the
DFT results so we start with a discussion of this picture.
Consider a graphene bilayer with an interlayer separation $d$
under an external electric field $E_{z,ext}$ along the $z$ direction.
Neglecting the finite thickness and crystalline inhomogeneity of the
graphene layers, and screening external to the bilayer, the Poisson equation is
\begin{equation}
\label{eq:poisson}
{\bm \nabla}\cdot{\bm E}=4\pi(-e) \left[n_1\delta(z)+n_2\delta(z-d)\right],
\end{equation}
where $n_1$ and $n_2$ are the net charge densities on the bottom and top layers, respectively.
If the bilayer is placed on a gate dielectric such as silicon dioxide (${\rm SiO_2}$)
and a voltage is applied between a gate and the bilayer, an excess charge carrier density
$n=n_1+n_2$ is supplied to the bilayer graphene and
redistributed between the top and bottom layers due to an external electric field.

In order to compare with our DFT calculations, we focus here on the
isolated bilayer case illustrated in Fig.\ref{fig:device_gap}, in which
the total excess density $n=n_1+n_2=0$.  Let us define $\delta n=n_2=-n_1$. From
Eq. (\ref{eq:poisson}), we obtain the screened electric field $E_z$ between the graphene
sheets of the bilayer to be
\begin{equation}
E_{z}-E_{z,ext}=4\pi e\delta n.
\end{equation}
Adding the corresponding Hartree potential to the external potential, we obtain
the screened interlayer potential difference as
\begin{equation}
\label{eq:Hartree_no_lattice}
U=U_{ext}+4\pi e^2 d \, \delta n ,
\end{equation}
where $U=e E_z d$ and $U_{ext} = e E_{z,ext} d$.

To estimate the relationship between $U$ and $U_{ext}$, we need
only a theory for the dependence of $\delta n$ on $U$.
In the $\pi$-orbital tight-binding model, $\delta n$ is given by the following integral over
the BZ:
\begin{equation}
\delta n = \sum_{i\in occ} 2\int_{BZ} {d^2 k \over (2\pi)^2} \big<\psi_i({\bm k})\big|{\sigma_z \over 2}\big|\psi_i({\bm k})\big> ,
\end{equation}
where $|\psi_i({\bm k})\rangle$ is a band eigenstate in the presence of $U$,
$\sigma_z={\rm diag(1,1,-1,-1)}$ in the (top,bottom)$\times$($A,B$) basis,
and the index $i$ runs over all occupied states.
The factor of 2 was included to account for spin degeneracy.

\begin{figure}[h]
\scalebox{0.45}{\includegraphics{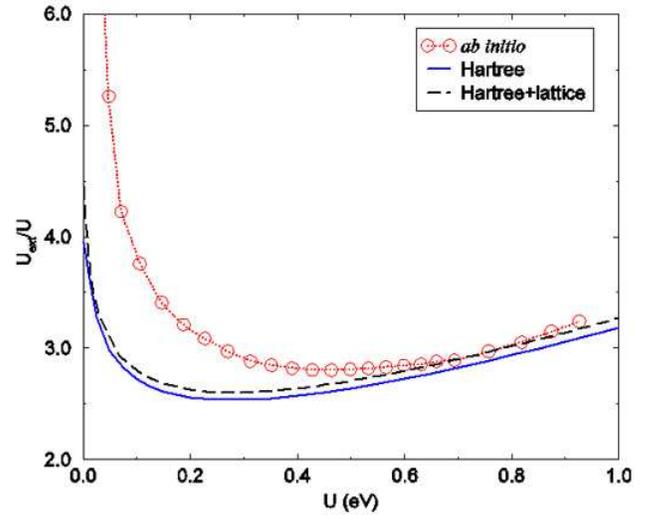}}
\caption{(Color online) The ratio of the external electric potential $U_{ext}$
to the interlayer energy difference inferred from the \textit{ab initio} DFT calculation
compared with the value of the same ratio in tight-binding model self-consistent Hartree
calculations, both with and without crystalline inhomogeneity corrections.
The tight-binding model calculations used $\gamma_0$=2.6 eV for the intralayer
tunneling amplitude, $\gamma_1$=0.34 eV for the interlayer tunneling amplitude,
and $\gamma_3$=0.3 eV for the interlayer $A$-$\tilde{B}$ coupling.
}
\label{fig:U_e}
\end{figure}

Figure \ref{fig:U_e} compares the screening ratio $U_{ext}/U$ obtained
from the \textit{ab initio} DFT calculations with the screening ratio from $\pi$-orbital tight-binding model
self-consistent Hartree
calculations with and without corrections that account for the crystalline inhomogeneity
within each layer as explained later.
The agreement between the three different approaches is generally good
especially at large potentials.

Note that as $U$ approaches zero, $U_{ext}/U$ increases in all approximations.
This property reflects increased screening as the gap decreases and is explained most
succinctly using the two-band continuum model\cite{mccann2006a} for the lowest-energy bands:
\begin{eqnarray}
H_{eff}&\approx&{U\over 2}
\left(
\begin{array}{cc}
1 & 0 \\
0 & -1 \\
\end{array}
\right)
-{1\over 2m}
\left(
\begin{array}{cc}
0 & (\pi^{\dagger})^2 \\
\pi^2 & 0 \\
\end{array}
\right)\nonumber \\
&=&{\bm a}\cdot{\bm \sigma} ,
\end{eqnarray}
where $\pi=p_x+i p_y$, $m={\gamma_1 \over 2v^2}$, ${\bm a}=\left(-{p_x^2-p_y^2 \over 2m},-{p_x p_y \over m},{U\over 2}\right)$,
and ${\bm \sigma}$ are $2\times 2$ Pauli matrices describing the top and bottom layer low-energy sites.
This Hamiltonian has simple spectra $\epsilon_{\pm}=\pm |{\bm a}|$
with eigenfunctions given by
\begin{equation}
\left|+\right>=
\left(
\begin{array}{c}
{\cos{\theta \over 2}} e^{-i\phi/2}\\
{\sin{\theta \over 2}} e^{i\phi/2}\\
\end{array}
\right),
\left|-\right>=
\left(
\begin{array}{c}
{-\sin{\theta \over 2}} e^{-i\phi/2} \\
{\cos{\theta \over 2}} e^{i\phi/2}\\
\end{array}
\right),
\end{equation}
where $\tan\theta={\sqrt{a_1^2+a_2^2}\over a_3}$ and $\tan\phi={a_2\over a_1}$.
It follows that
\begin{eqnarray}
\label{eq:deltan}
\delta n &=& 4 \int_{|p|<p_c} {d^2 p \over (2\pi\hbar)^2} \big<-,{\bm p}\big|{\sigma_z \over 2}\big|-,{\bm p}\big> \nonumber \\
&=&-{1\over \pi\hbar^2}\int_0^{p_c}pdp \cos\theta({\bm p}) \nonumber \\
&=&-{m U \over 2\pi \hbar^2} \ln\left[x_c+\sqrt{x_c^2+1}\right] ,
\end{eqnarray}
where $x_c={p_c^2 \over m U}$.  We have inserted a factor of 4 in this
continuum model calculation to
account for both spin($\uparrow$ and $\downarrow$) and valley($K$ and $K'$) degeneracies.
The integral over wave vector was cut off at the radius $p_c\sim \sqrt{2m \gamma_1}$
beyond which the continuum model fails.

Inserting Eq. (\ref{eq:deltan}) in Eq. (\ref{eq:Hartree_no_lattice}), we obtain
\begin{eqnarray}
{U_{ext}\over U}&=&1-4\pi e^2 d\, {\delta n \over U} \\
&=&1+2\left(d\over a_B\right)\left(m\over m_e\right) \ln\left[x_c+\sqrt{x_c^2+1}\right] \nonumber ,
\end{eqnarray}
where $a_B=\hbar^2/m_e e^2$ is the Bohr radius and $m_e$ is the bare electron mass.
For small $U$, $x_c$ is large and this simplifies to
\begin{equation}
{U_{ext}\over U}\approx 2\left(d\over a_B\right)\left(m\over m_e\right) \ln\left[{2 p_c^2 \over m U}\right].
\end{equation}
A related observation concerning the logarithmic divergence of the screening ratio at small
gate voltages was made previously by McCann.\cite{mccann2006b}
All three of our calculations exhibit this increased screening at weak external potentials, with the
largest upturn in the \textit{ab initio} calculations.

\subsection{Lattice Hartree potential models}
We now turn our attention to one important
contribution to discrepancies between the \textit{ab initio} DFT results and the
predictions of self-consistent Hartree
models similar to those described above, the role of
crystalline inhomogeneity in bilayer and single-layer graphene electrostatics.
We consider a general two-body interaction term $\hat{V}$,
\begin{equation}
\label{eq:two_body_operator}
\hat{V}={1 \over 2}\sum_{\lambda_1',\lambda_2',\lambda_1,\lambda_2}\left<\lambda_1' \lambda_2'\right| V \left|\lambda_1 \lambda_2 \right>c_{\lambda_1'}^{\dagger} c_{\lambda_2'}^{\dagger}c_{\lambda_2} c_{\lambda_1},
\end{equation}
where $c_{\lambda}^{\dagger}$ and $c_{\lambda}$ are creation and annihilation
operators for a state $\lambda$.  To capture the main consequences of crystalline inhomogeneity, we
assume that the $\pi$-orbital Bloch states with crystal momentum ${\bm k}$ can be written as a linear combination of
atomic orbitals,
\begin{equation}
\psi_{{\bm k},\lambda}({\bm x})={1\over \sqrt{N}}\sum_{{\bm R}} \; e^{i{\bm k}\cdot{\bm R}}\phi_{\lambda}({\bm x}-{\bm R}-{\bm \tau}_{\lambda})
\end{equation}
where $\phi_{\lambda}$ is an atomiclike $\pi$ orbital, ${\bm R}$ is a lattice vector, ${\bm \tau}_{\lambda}$
is the displacement of the sites in a unit cell with respect to the lattice vector, and $N$ is the number of lattice sites.
If we assume that the overlap of $\phi_{\lambda}$-orbitals centered on different sites can be neglected and
ignore the $\hat z$ direction spread of the graphene sheets, the interaction Hamiltonian simplifies to
\begin{equation}
\label{eq:two_body_operator_lattice}
\hat{V}= {1 \over 2 \Omega}\sum_{{\bm k_1},{\bm k_2},{\bm q}}\sum_{\lambda_1,\lambda_2}\, \tilde{V}_{\lambda_1,\lambda_2}({\bm q}) \; c_{{\bm k_1}+{\bm q},\lambda_1}^{\dagger} c_{{\bm k_2}-{\bm q},\lambda_2}^{\dagger} c_{{\bm k_2},\lambda_2} c_{{\bm k_1},\lambda_1},
\end{equation}
where $\Omega$ is the area of the two-dimensional plane,
\begin{equation}
\label{eq:2Dpot}
\tilde{V}_{\lambda_1,\lambda_2}({\bm q})=V_{\lambda_1,\lambda_2}({\bm q}) \; w_{\lambda_1}(-{\bm q}) w_{\lambda_2}({\bm q}) \;  e^{i {\bm q}\cdot({\bm \tau}_{\lambda_1}-{\bm \tau}_{\lambda_2})},
\end{equation}
\begin{equation}
\label{eq:2Dpot_free}
V_{\lambda_1,\lambda_2}({\bm q})=\int d{\bm x} \; e^{-i{\bm q}\cdot{\bm x}} \; V_{\lambda_1,\lambda_2}({\bm x}),
\end{equation}
and
\begin{equation}
w_{\lambda}({\bm q})=\int d{\bm x} \;  e^{-i{\bm q}\cdot{\bm x}}\; \left|\phi_{\lambda}({\bm x})\right|^2.
\end{equation}
Note that the labels ${\bm k_1}$ and ${\bm k_2}$ are restricted to the BZ, while ${\bm q}$ runs over the two-dimensional plane.
In Eq. (\ref{eq:2Dpot_free}), $V_{\lambda_1,\lambda_2}({\bm x}) = e^2/|{\bm x}|$ when $\lambda_1$ and $\lambda_2$ refer to sites in the
same layer and $V_{\lambda_1,\lambda_2}({\bm x}) = e^2/\sqrt{|{\bm x}|^2+d^2}$ when $\lambda_1$ and $\lambda_2$ refer
to layers separated by $d$.  It follows that $V_{\lambda_1,\lambda_2}({\bm q}) = 2\pi e^2/|{\bm q}|$ for labels
in the same layer and $V_{\lambda_1,\lambda_2}({\bm q}) = 2 \pi e^2 \exp(-|{\bm q}|d)/|{\bm q}|$ for labels
in different layers.  Since the total charge of the bilayer is fixed in our calculations, only the differences
between the various $\tilde{V}_{\lambda_1,\lambda_2}({\bm q})$ values are relevant.
For explicit calculations, we have used a Gaussian form factor $w_{\lambda}({\bm q})=e^{-|{\bm q}|^2 r_0^2/2}$
corresponding to $\left|\phi_{\lambda}({\bm x})\right|^2\propto e^{-|{\bm x}|^2/2r_0^2}$,
where $r_0\sim 0.48$ $\rm \AA$ was obtained by fitting to the DFT valence orbitals.

This two-body Hamiltonian can be used to account for crystalline inhomogeneity
in a graphene bilayer system with arbitrary electronic correlations.  To compare with
the \textit{ab initio} DFT calculations, we consider interactions in a mean-field
Hartree approximation in which the interaction contribution to the single-particle
Hamiltonian is
\begin{equation}
\hat{V}^{(H)}= \sum_{{\bm k},a,\sigma} \epsilon_{a\sigma}^{(H)} c_{{\bm k},a\sigma}^{\dagger} c_{{\bm k},a\sigma}
\end{equation}
where $a$ and $\sigma$ denote layer and sublattice degrees of freedom.
Here,
\begin{equation}
\epsilon_{a\sigma}^{(H)}= \sum_{a',\sigma'} \tilde{V}_{a\sigma,a'\sigma'} n_{a'\sigma'},
\end{equation}
where $n_{a\sigma}={2\over \Omega} \sum_{{\bm k}} \left<c_{{\bm k},a\sigma}^{\dagger} c_{{\bm k},a\sigma}\right>$
including spin degeneracy and
\begin{equation}
\tilde{V}_{a\sigma,a'\sigma'} = \sum_{{\bm G}}\tilde{V}_{a\sigma,a'\sigma'}({\bm G}),
\end{equation}
with ${\bm G}$ a triangular lattice reciprocal-lattice vector.


As explained in the Introduction, interlayer tunneling in graphene
leads to high-energy bands which favor the $\tilde{A}$-$B$ sites
and low-energy bands that favor the $A$-$\tilde{B}$ sites.
Since the low-energy bands respond most strongly to the external potential,
we can expect that the charge transfer occurs
more strongly on the $A$-$\tilde{B}$ sites, and that the
screening potential should be larger on these sites.
Instead of a single-interlayer Hartree screening potential,
two Hartree potentials for low and high bands must be calculated separately:
\begin{eqnarray}
\epsilon_{l}^{(H)}&=&\epsilon_{\tilde{B}}^{(H)}-\epsilon_{A}^{(H)}, \\
\epsilon_{h}^{(H)}&=&\epsilon_{\tilde{A}}^{(H)}-\epsilon_{B}^{(H)}. \nonumber
\end{eqnarray}
When only the $\bm{G}=0$ term is retained in the reciprocal-lattice vector sum,
\begin{equation}
\epsilon_{l}^{(H0)}=\epsilon_{h}^{(H0)}=2\pi e^2 d\; (n_{\tilde{A}}+n_{\tilde{B}}-n_A-n_B),
\end{equation}
and Eq. (\ref{eq:Hartree_no_lattice}) is recovered.
\begin{figure}[h]
\scalebox{0.45}{\includegraphics{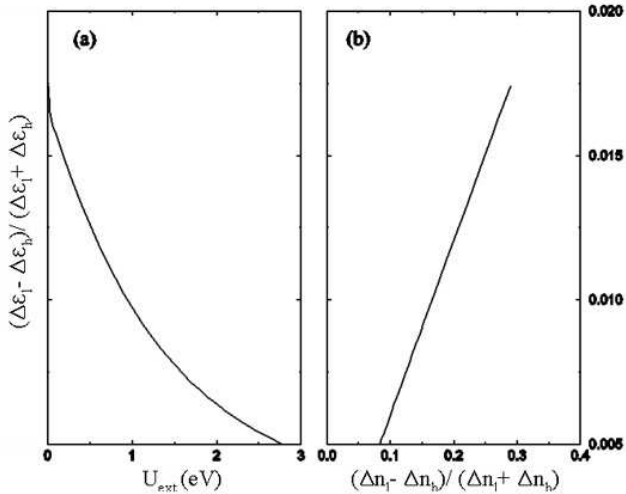}}
\caption{Splitting of Hartree potentials
$\big(\epsilon_{l}^{(H)}-\epsilon_{h}^{(H)}\big)/\big(\epsilon_{l}^{(H)}+\epsilon_{h}^{(H)}\big)$
as a function of (a) the external electric potential $U_{ext}$
and (b) the corresponding density inhomogeneity $\big( \Delta n_l-\Delta  n_h\big)/ \big(\Delta n_{l}+\Delta n_{h}\big)$
in the lattice Hartree potential model.}
\label{fig:U0_H_N_sp}
\end{figure}
It turns out that the sum over reciprocal-lattice vectors can be
truncated with good accuracy at the first shell.  Noting that
$e^{-|{\bm G}|d}\ll 1$, we find for the crystalline inhomogeneity corrections
\begin{eqnarray}
\label{eq:H_splitting}
\epsilon_{l}^{(H1)}&\approx& 2\pi e^2 d \, \alpha({\bm G}) ( 6\Delta n_l-3\Delta n_h ), \\
\epsilon_{h}^{(H1)}&\approx& 2\pi e^2 d \, \alpha({\bm G}) ( 6\Delta n_h-3\Delta n_l ), \nonumber
\end{eqnarray}
where $\Delta n_l=n_{\tilde{B}}-n_A$, $\Delta n_h=n_{\tilde{A}}-n_B$,
and $\alpha({\bm G})=e^{-|{\bm G}|^2 r_0^2}/ |{\bm G}|d \approx 0.0136$.
Thus, the inhomogeneity effect results in more screening as expected, but
as indicated by the black dashed line in Fig. \ref{fig:U_e}, it
is not able to account for the largest part of the discrepancy between
DFT and model results.

As the external electric potential is decreased, the difference between
low-energy and high-energy site occupancies is increased.
The difference in Hartree potentials rises correspondingly, as illustrated
in Fig. \ref{fig:U0_H_N_sp}(a). From Eq. (\ref{eq:H_splitting}), we can estimate the relation between
the splitting of the Hartree potentials and the density inhomogeneity:
\begin{equation}
{\epsilon_{l}^{(H)}-\epsilon_{h}^{(H)}\over \epsilon_{l}^{(H)}+\epsilon_{h}^{(H)}} \approx {9\over 2}\alpha({\bm G}) {\Delta n_{l}-\Delta n_{h}\over \Delta n_{l}+\Delta n_{h}},
\end{equation}
where the coefficient ${9\over 2}\alpha({\bm G})$ is given by $\sim$0.0612 [Fig. \ref{fig:U0_H_N_sp}(b)].

\section{Discussion}
Our DFT calculations of external potential induced gaps in the electronic structure of
graphene bilayers confirm the simple picture provided by phenomenological tight-binding models.
The \textit{ab initio} calculations include a number of effects not contained in the model
calculations.  For example, the occupied $\sigma$ orbitals within each graphene plane, which
are neglected in the $\pi$-orbital tight-binding model, will be slightly
polarized by the external electric field and contribute to screening.  In the DFT calculations,
not only Hartree potentials but also exchange-correlation potentials will be altered by
an external electric field and influence the screening process.  Since the exchange potential is
attractive, its contribution to the total potential will lower energies in a layer more
as the density is increased.  The exchange potential therefore makes a negative contribution to the screening ratio.
The quantitative discrepancies between the DFT and phenomenological model reflect the combination of these
and other additional effects contained in the DFT calculations, and strong sensitivity to
intralayer and interlayer tunneling amplitudes which may not be evaluated with perfect accuracy
by DFT.  We also note that the low-energy eigenstates in bilayer graphene are coherent combinations
of amplitudes on both layers, which implies that interlayer exchange interactions will be
substantial.  This kind of effect is absent in the exchange-correlation potentials commonly used in DFT.
Indeed, it is entirely possible that DFT calculations do not predict accurate values for the
screening ratio.  We believe that there is strong motivation for capacitive studies of the
interlayer screening properties of graphene bilayers using an experimental arrangement similar to
that in Fig. \ref{fig:device_gap}.

In summary, we have used \textit{ab initio} density functional theory calculations
to study the gate-voltage tunable gap in the electronic structure of bilayer graphene.
The electric-field dependence of the on-site energy difference and the interlayer
tunneling amplitude were extracted from the DFT calculation results by
fitting to tight-binding model expressions for high-symmetry point graphene bilayer
band eigenvalues.  The screening effect seen in the DFT calculations can be explained by
a tight-binding model self-consistent Hartree method including crystalline inhomogeneity corrections,
although the DFT screening is stronger especially for weak external potentials.

\acknowledgements

This work was supported
by the Welch Foundation (Houston, TX) under Grant No. F-1473 and No. F-0934, by the Texas Advanced Computing Center (TACC),
University of Texas at Austin, by Seagate Corporation, and by the Department of Energy under Grant No. DE-FG03-96ER45598.
B.S. and S.K.B. thank SRC-NRI (SWAN) for financial support.
The authors gratefully acknowledge helpful conversations with A. Castro-Neto and V. Fal'ko.


\begin{thebibliography}{999}
\bibitem{novoselov2004} K. S. Novoselov, A. K. Geim, S. V. Morozov, D. Jiang, Y. Zhang, S.V.Dubonos, I. V. Grigorieva, and A. A. Firsov, Science {\bf 306}, 666 (2004).
\bibitem{novoselov2005a} K. S. Novoselov, D. Jiang, F. Schedin, T. J. Booth, V. V. Khotkevich, S. V. Morozov, and A. K. Geim, Proc. Natl Acad. Sci. USA {\bf 102}, 10451 (2005).
\bibitem{zhang2005a} Y. Zhang, Joshua P. Small, Michael E. S. Amori, and Philip Kim, Phys. Rev. Lett. {\bf 94}, 176803 (2005).
\bibitem{novoselov2006} K. S. Novoselov, E. McCann, S. V. Morozov, V.I.Fal'ko, M. I. Katsnelson, U. Zeitler, D. Jiang, F. Schedin, and A. K. Geim, Nat. Phys. {\bf 2}, 177 (2006).
\bibitem{ohta2006} Taisuke Ohta, Aaron Bostwick, Thomas Seyller, Karsten Horn, and Eli Rotenberg, Science {\bf 313}, 951 (2006).
\bibitem{novoselov2005b} K. S. Novoselov, A. K. Geim, S. V. Morozov, D. Jiang, M. I. Katsnelson, I. V. Grigorieva, S. V. Dubonos, and A. A. Firsov, Nature {\bf 438}, 197 (2005).
\bibitem{zhang2005b} Y. Zhang, Yan-Wen Tan, Horst L. Stormer, and Philip Kim, Nature {\bf 438}, 201 (2005).
\bibitem{kane2005}C. L. Kane and E. J. Mele, Phys. Rev. Lett. {\bf 95}, 226801 (2005); C. L. Kane and E. J. Mele, Phys. Rev. Lett. {\bf 95}, 146802 (2005).
\bibitem{sinitsyn2006} N. A. Sinitsyn, J. E. Hill, Hongki Min, Jairo Sinova, and A. H. MacDonald, Phys. Rev. Lett. {\bf 97}, 106804 (2006); Hongki Min, J. E. Hill, N. A. Sinitsyn, B. R. Sahu, Leonard Kleinman, and A. H. MacDonald, Phys. Rev. B {\bf 74}, 165310 (2006).
\bibitem{latil2006} Sylvain Latil and Luc Henrard, Phys. Rev. Lett. {\bf 97}, 036803 (2006); B. Partoens and F. M. Peeters, Phys. Rev. B {\bf 74}, 075404 (2006).
\bibitem{mccann2006a} Edward McCann and Vladimir I. Fal'ko, Phys. Rev. Lett. {\bf 96}, 086805 (2006).
\bibitem{mccann2006b} Edward McCann, Phys. Rev. B {\bf 74}, 161403(R) (2006).
\bibitem{castro2006} Eduardo V.Castro, K.S.Novoselov, S. V. Morozov, N. M. R. Peres, J. M. B. Lopes dos Santos, Johan Nilsson, F. Guinea, A. K. Geim, and A. H. Castro Neto, cond-mat/0611342 (unpublished).
\bibitem{kohn1965} W. Kohn and L. J.Sham, Phys. Rev. {\bf 140}, A1133 (1965).
\bibitem{blaha2001} P. Blaha, K. Schwarz, G. K. H. Madsen, D. Kvasnicka, and J. Luitz, \textit{WIEN2K, An Augmented Plane Wave+Local Orbitals Program for Calculating Crystal Properties} (K.Schwarz, Techn. Universit\"{a}t Wien, Vienna, Austria, 2001).
\bibitem{perdew1996} J. P. Perdew, K. Burke, and M. Ernzerhof, Phys. Rev. Lett. {\bf 77}, 3865 (1996).
\bibitem{stahn2001} J. Stahn, U. Pietsch, P. Blaha, and K. Schwarz, Phys. Rev. B.{\bf 63}, 165205 (2001).
\end{thebibliography}
\end{document}